\documentclass[prl,unsortedaddress,showpacs,amssymb,amsmath,
superscriptaddress, preprint]{revtex4}
\usepackage{epsf,graphicx}
\begin{document}
\newcommand{\deltl}{$\Delta L = 2$}
\newcommand{\pmmdec}{$\Xi^- \to p \mu^- \mu^-$}
\newcommand{\smgauge}{$SU(2)_L \times U(1)_Y$}
\newcommand{\casdec}{$\Xi^- \to \Lambda \pi^-$}
\newcommand{\lamdec}{$\Lambda \to p \pi^-$}
\newcommand{\kdec}{$K^- \to \pi^-\pi^-\pi^+$}
\newcommand{\ktpi}{$K^{\pm} \to \pi^{\mp} \pi^{\mp} \pi^{\pm}$}
\title {
Search for the Lepton-Number-Violating Decay \pmmdec}

\date{\today}

\affiliation{Institute of Physics, Academia Sinica, Taipei 11529, Taiwan,
             Republic of China}
\affiliation{University of California, Berkeley, California 94720, USA}
\affiliation{Fermi National Accelerator Laboratory, Batavia, Illinois
             60510, USA}
\affiliation{Universidad de Guanajuato, 37000 Le\'{o}n, Mexico}
\affiliation{Illinois Institute of Technology, Chicago, Illinois 60616, USA}
\affiliation{Universit\'{e} de Lausanne, CH-1015  Lausanne, Switzerland}
\affiliation{Lawrence Berkeley National Laboratory, Berkeley, California
             94720, USA}
\affiliation{University of Michigan, Ann Arbor, Michigan 48109, USA}
\affiliation{University of South Alabama, Mobile, Alabama 36688, USA}
\affiliation{University of Virginia, Charlottesville, Virginia 22904, USA}

\author{D.~Rajaram}
  \altaffiliation[Present address: ]{University of Michigan, Ann Arbor, MI 48109, USA.}
  \affiliation{Illinois Institute of Technology, Chicago, Illinois 60616, USA}
\author{R.~A.~Burnstein}
  \affiliation{Illinois Institute of Technology, Chicago, Illinois 60616, USA}
\author{A.~Chakravorty}
  \affiliation{Illinois Institute of Technology, Chicago, Illinois 60616, USA}
\author{A.~Chan}
  \affiliation{Institute of Physics, Academia Sinica, Taipei 11529, Taiwan,
               Republic of China}
\author{Y.~C.~Chen}
  \affiliation{Institute of Physics, Academia Sinica, Taipei 11529, Taiwan,
               Republic of China}
\author{W.~S.~Choong}
  \affiliation{University of California, Berkeley, California 94720, USA}
  \affiliation{Lawrence Berkeley National Laboratory, Berkeley, California
               94720, USA}
\author{K.~Clark}
  \affiliation{University of South Alabama, Mobile, Alabama 36688, USA}
\author{E.~C.~Dukes}
  \affiliation{University of Virginia, Charlottesville, Virginia 22904, USA}
\author{C.~Durandet}
  \affiliation{University of Virginia, Charlottesville, Virginia 22904, USA}
\author{J.~Felix}
  \affiliation{Universidad de Guanajuato, 37000 Le\'{o}n, Mexico}
\author{G.~Gidal}
  \affiliation{Lawrence Berkeley National Laboratory, Berkeley, California
               94720, USA}
\author{P.~Gu}
  \affiliation{Lawrence Berkeley National Laboratory, Berkeley, California
               94720, USA}
\author{H.~R.~Gustafson}
  \affiliation{University of Michigan, Ann Arbor, Michigan 48109, USA}
\author{C.~Ho}
  \affiliation{Institute of Physics, Academia Sinica, Taipei 11529, Taiwan,
               Republic of China}
\author{T.~Holmstrom}
  \affiliation{University of Virginia, Charlottesville, Virginia 22904, USA}
\author{M.~Huang}
  \affiliation{University of Virginia, Charlottesville, Virginia 22904, USA}
\author{C.~James}
  \affiliation{Fermi National Accelerator Laboratory, Batavia, Illinois
               60510, USA}
\author{C.~M.~Jenkins}
  \affiliation{University of South Alabama, Mobile, Alabama 36688, USA}
\author{D.~M.~Kaplan}
  \email[Corresponding author.~ Electronic address:  ]
        {kaplan@iit.edu}
  \affiliation{Illinois Institute of Technology, Chicago, Illinois 60616, USA}
\author{L.~M.~Lederman}
  \affiliation{Illinois Institute of Technology, Chicago, Illinois 60616, USA}
\author{N.~Leros}
  \affiliation{Universit\'{e} de Lausanne, CH-1015 Lausanne, Switzerland}
\author{M.~J.~Longo}
  \affiliation{University of Michigan, Ann Arbor, Michigan 48109, USA}
\author{F.~Lopez}
  \affiliation{University of Michigan, Ann Arbor, Michigan 48109, USA}
\author{L.~C.~Lu}
  \affiliation{University of Virginia, Charlottesville, Virginia 22904, USA}
\author{W.~Luebke}
  \affiliation{Illinois Institute of Technology, Chicago, Illinois 60616, USA}
\author{K.~B.~Luk}
  \affiliation{University of California, Berkeley, California 94720, USA}
  \affiliation{Lawrence Berkeley National Laboratory, Berkeley, California
               94720, USA}
\author{K.~S.~Nelson}
  \affiliation{University of Virginia, Charlottesville, Virginia 22904, USA}
\author{H.~K.~Park}
  \affiliation{University of Michigan, Ann Arbor, Michigan 48109, USA}
\author{J.-P.~Perroud}
  \affiliation{Universit\'{e} de Lausanne, CH-1015 Lausanne, Switzerland}
\author{H.~A.~Rubin}
  \affiliation{Illinois Institute of Technology, Chicago, Illinois 60616, USA}
\author{P.~K.~Teng}
  \affiliation{Institute of Physics, Academia Sinica, Taipei 11529, Taiwan,
               Republic of China}
\author{J.~Volk}
  \affiliation{Fermi National Accelerator Laboratory, Batavia, Illinois
               60510, USA}
\author{C.~G.~White}
  \affiliation{Illinois Institute of Technology, Chicago, Illinois 60616, USA}
\author{S.~L.~White}
  \affiliation{Illinois Institute of Technology, Chicago, Illinois 60616, USA}
\author{P.~Zyla}
  \affiliation{Lawrence Berkeley National Laboratory, Berkeley, California
               94720, USA}
\collaboration{HyperCP Collaboration}
\noaffiliation

\begin{abstract}
A sensitive search for the lepton-number-violating decay $\Xi^-\to p \mu^-\mu^-$ has been performed
using a sample of $\sim10^9$ $\Xi^-$ hyperons produced in 800\,GeV/$c$ $p$-Cu collisions. 
We obtain $\mathcal{B}(\Xi^-\to p \mu^-\mu^-)< 4.0\times 10^{-8}$ at 90\% confidence, improving on the best  previous limit by four orders of magnitude.
\end{abstract}


\maketitle

The conservation of lepton and lepton-family numbers (the latter notably violated by neutrino mixing) is one of the fundamental
puzzles in physics~\cite{thbackground}. These conservation laws, while satisfied in the standard model, are based on no known deeper principle. For example, lepton-number conservation, unlike
such global conservation laws as that for electric charge, is not associated
with any local gauge invariance. 
With neutrino oscillation now established experimentally~\cite{neutosc}, models incorporating neutrino mass and lepton-family-number nonconservation must be considered. These typically feature~\cite{lnv-models} Majorana neutrinos, whose exchange changes the
total lepton number $L$ by two units. Conversely, the observation of \deltl\ processes could imply the existence of massive Majorana neutrinos~\cite{lnv-maj}. Experimental searches for
lepton-number nonconservation are thus of fundamental importance.

\begin{figure}
\centerline {\includegraphics[width=1.8 in]{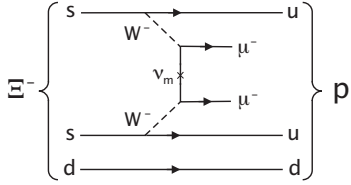}}
\caption[A possible diagram for the decay \pmmdec]{A possible diagram for
the decay \pmmdec; $\nu_m$ represents a Majorana neutrino. }
\label{quarkdiag}
\end{figure}

The most stringent 
limits on lepton-number nonconservation come from searches 
for neutrinoless double-beta ($0\nu\beta\beta$) decay, the best 
 being
$t_{1/2} > 1.9 \times 10^{25}$\,y (C.L.=90\%) for 
$^{76}$Ge~\cite{dbdec:heidelmoscow}. However, searches for $H_1 \to H_2 \mu \mu$, where $H_{1,2}$
are hadrons, can provide complementary information. 
Figure~\ref{quarkdiag} shows a possible mechanism. 
Predicting rates for such decays is challenging, since
they depend sensitively on details of the underlying dynamics of 
neutrino mixing and of the hadronic matrix element~\cite{littenberg2,shrock}. These rates are unconstrained by limits on $0\nu\beta\beta$ decays 
and by
conversion rates of muons to electrons in nuclear interactions~\cite{littenberg}. 
Limits have been set on such processes in $D$, $B$, and $K$ decays at branching-ratio sensitivities ranging 	 
from $10^{-4}$ to $10^{-9}$~\cite{PDG}. However, 
in the baryon sector, experimental knowledge on such decays
is meager. The only available limits are $\mathcal{B}(\Lambda^+_c\to\Sigma^-\mu^+\mu^+)<7.0\times10^{-4}$~\cite{PDG} and $\mathcal{B}(\Xi^-\to p\mu^-\mu^-)<3.7 \times 10^{-4}$~\cite{littenberg}; the latter, based on 8150 $\Xi^-$ events observed with the Brookhaven
National Laboratory 31\,in bubble chamber~\cite{bnl74}, remains until now the
best  limit on \deltl\ processes in the hyperon sector.

\begin{figure}
\centerline {\includegraphics[width=3.4 in]{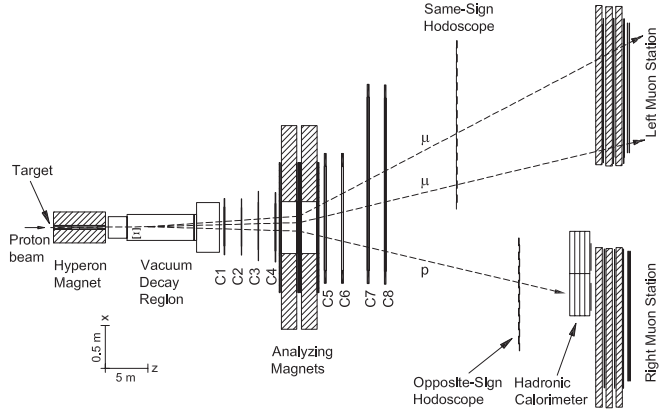}}
\caption{Plan view of the HyperCP spectrometer. C1--C8 are multiwire proportional chambers. Note that the $z$ scale is compressed by a factor of 10 compared to the $x$ scale. 
\label{fig:e871}}
\end{figure}

We report a search for \pmmdec\ in the HyperCP experiment (Fermilab E871)
based on $\sim 10^9$ $\Xi^-$ decays --- a vastly larger sample than previously available. 
The experimental arrangement (Fig.~\ref{fig:e871}) and muon selection criteria are described in \cite{hypercp-kmumu}.
In brief, a negatively charged secondary beam was formed by the interaction of 800\,GeV/$c$ primary protons from the Tevatron in a $0.2\times 0.2\times 6\,{\rm cm}^3$ copper target, with the sign and momenta of secondaries selected by a 6.096-m-long curved collimator within a 1.667\,T dipole magnetic field.  
The mean momentum of the secondaries was about 160\,GeV/$c$, with $\approx25$\% FWHM momentum spread. The typical secondary-beam rate was 13\,MHz at the exit of the collimator. Hyperon decays occurring within a 13-m-long evacuated 
 pipe (the ``vacuum decay region" of Fig.~\ref{fig:e871}) were reconstructed in three dimensions 
 in a series of high-rate multiwire proportional chambers (C1--C8),  with wire spacings increasing
 from 1 to 2\,mm. A pair of dipole magnets (``analyzing magnets") deflected charged particles
 horizontally with a transverse-momentum kick of 1.43\,GeV/$c$. A pair of muon detector stations
 consisted of planes of vertical and horizontal proportional tubes with 2.54\,cm pitch, interspersed with three layers of $\approx$0.75-m-thick iron absorber, followed by vertical and horizontal scintillation hodoscopes.

The trigger for online data acquisition used two scintillation-counter hodoscopes (``same-sign" and ``opposite-sign hodoscopes" in Fig.~\ref{fig:e871}), located
sufficiently far downstream of the analyzing magnets that the
hyperon decay products were well separated from
the secondary beam. At least one hodoscope hit from a negative (same-sign) track in coincidence with one from a positive (opposite-sign) track was required for a trigger. To suppress muon and low-energy backgrounds,
the trigger also required a minimum energy deposit in the
hadronic calorimeter. The calorimeter energy threshold was set sufficiently low that the
calorimeter trigger was more than $99$\% efficient for protons from  $\Xi^-$ decays within the secondary-beam momentum range.

\begin{figure}
\vspace{-.4in}
\centerline{\includegraphics[width=2.6 in]{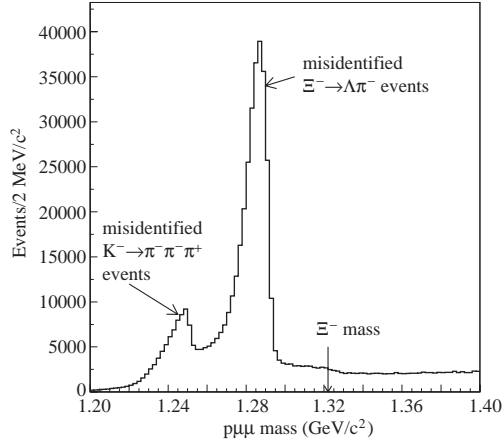}}
\caption{$p\mu^-\mu^-$ invariant-mass distribution before full application of
selection
requirements; note broadened and shifted peaks due to $\pi$--$\mu$ misidentification.}
\label{fig:raw}
\end{figure}

The  \pmmdec\ decay gives two like-sign muon
tracks and a proton track originating from a common vertex.  The first stage of
data reduction selected events with three tracks,  at least one being a muon
track. A muon track was one with hits in at least two of three
muon proportional-tube planes in both the $x$ and $y$ views. Figure~\ref{fig:raw} shows the $p\mu^-\mu^-$ invariant-mass distribution, which is entirely dominated by background due to misidentified pions. In the next analysis stage, 
the muon requirement was tightened by requiring in-time hits
in the muon hodoscopes corresponding to hits in the proportional tubes, and two negative muon
tracks were required as well as a third track of opposite sign (assumed to be the proton). The total momentum of the three tracks was required to be between
120 and 250\,GeV/$c$, consistent with the momentum spectrum of the
secondary beam. 

Next, requirements were imposed on the decay vertex, which was reconstructed by fitting the three tracks to a common vertex using only the 
hits in C1--C4. To reduce backgrounds from
interactions near $z = 0$ in the collimator material and 
windows, and in windows near $z=1330$\,cm, the reconstructed vertex was required to
lie well within the vacuum decay region ({i.e.}, between 65 and 1285\,cm downstream of
the end of the collimator).  To suppress the copious backgrounds due to two-vertex hyperon decay (such as the  \casdec,
\lamdec\ decay chain with both pions
misidentified as muons due to in-flight decay or
punch-through in the muon detectors), requirements were imposed on the $\chi^2$
of the single-vertex fit, as well as on the average distance in the $x$--$y$ plane
between pairs of tracks at the $z$ position of
the fitted vertex. (This average separation was calculated from the wire hits in C1--C4 without imposing the single-vertex constraint.) Based on the vertex-$\chi^2$ and average-separation distributions of clean  \kdec\  decays, candidates were accepted if
the $\chi^2/{\rm dof} $ was less than 2.5 and the average separation was
less than 0.2\,cm. 
To ensure that the parent hyperon was
produced in the target, the parent-particle trajectory was traced back to the
target and required to originate within $\pm$0.5\,cm of the
target center in both $x$ and $y$ (corresponding to $\pm$7 (5.6) standard deviations ($\sigma$) of the $x_{\rm target}$ $(y_{\rm target})$ resolution).

For further background rejection, events passing the above requirements were subjected to three
invariant-mass requirements ($K$, $\Lambda$, and $\Xi$ vetoes, respectively): 
(1) events with invariant mass between 473 and 513\,MeV/$c^2$ (corresponding to $\pm10\,\sigma$ of the $K^-$ mass resolution) under the $\pi^-\pi^-\pi^+$
hypothesis
were rejected as being \kdec\ decays; 
(2) if the $p\pi^-$ invariant mass for either pair of oppositely charged 
tracks was between 1100 and 1125\,MeV/$c^2$ ($\pm7\,\sigma$ of the resolution), the
event was rejected as having a  
$\Lambda$; 
(3) if the invariant mass under the $p\pi^-\pi^-$ hypothesis was between
1315 and 1330\,MeV/$c^2$  ($\pm4\,\sigma$ of the resolution) the event was rejected as being a \casdec, \lamdec\ 
decay. These requirements were based on the observed \casdec\ sample and
Monte Carlo (MC) simulations of the signal decay that indicated, for example, that $\Xi^-\to p \mu^- \mu^-$ decays interpreted as $\Xi^- \to p \pi^- \pi^-$ would yield a parent $\Xi^-$ mass exceeding 1330\,MeV/$c^2$.
In addition, since $K^- \rightarrow \pi^- \pi^- \pi^+$ decays
    on average give a lower value of positive-track 
    momentum than do $\Xi^- \rightarrow p \mu^- \mu^-$ decays,
    the positive-track momentum was required to exceed 56\%
    of the total three-track momentum.
    This requirement was estimated to be 91\% efficient for \pmmdec\  decays while rejecting over 90\% of \kdec\ decays.

\begin{figure}
\centerline {\includegraphics[width=2.5 in]{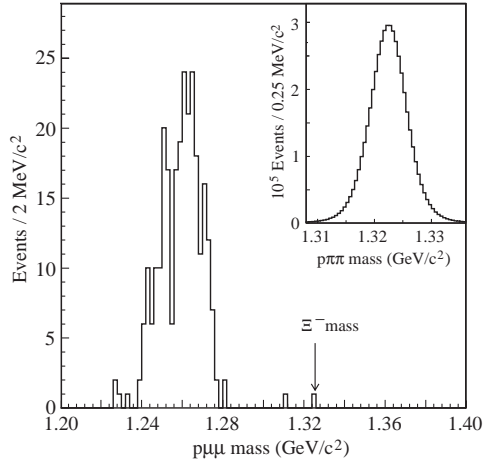}}
\caption{$p\mu^-\mu^-$ invariant-mass distribution after application of all
selection requirements; inset: $p\pi^-\pi^-$ invariant-mass distribution from the 
normalization sample.}
\label{pmumufinal}
\end{figure}

Events that passed all of the above requirements were reconstructed under
the $p \mu^- \mu^-$ hypothesis. Figure~\ref{pmumufinal} shows the resulting
invariant-mass distribution.
There is one event within the 11-MeV/$c^2$-wide search region corresponding to $\pm 3\,\sigma$ about the 1321.31\,MeV/$c^2$~\cite{PDG} $\Xi^-$ mass. We treat this event as background in determining an upper limit on $\mathcal{B}(\Xi^- \rightarrow p \mu^- \mu^-)$. Since there is one event in the 11-MeV/$c^2$-wide sideband below the search region and zero in that above the search region, the background is estimated as $0.5\pm0.5$~events.

The normalizing mode for this search was $\Xi^- \rightarrow \Lambda\pi^- \rightarrow p\pi^-\pi^-$, 
 recorded using the same trigger as the signal mode.
It was studied in a ``prescaled" sample: only every 100th event passing the first stage of data reduction  was fully analyzed. 
The selection requirements were the same as for the signal mode, except
that the $\Xi$- and $\Lambda$-veto, single-vertex, and muon requirements
were not made 
and the $\Lambda$-decay vertex was allowed to lie beyond the 
vacuum decay region. The resulting invariant-mass distribution is shown in Fig.~\ref{pmumufinal} (inset).
The total number of reconstructed $\Xi^- \rightarrow \Lambda\pi^- \rightarrow p\pi^-\pi^-$ events that would have passed these  requirements had every event been analyzed was $(4.92\pm0.30)
\times 10^8$,  the error arising from
 the uncertainty in the background subtraction.
 
\begin{figure}
\centerline{\includegraphics[height=2.35 in]{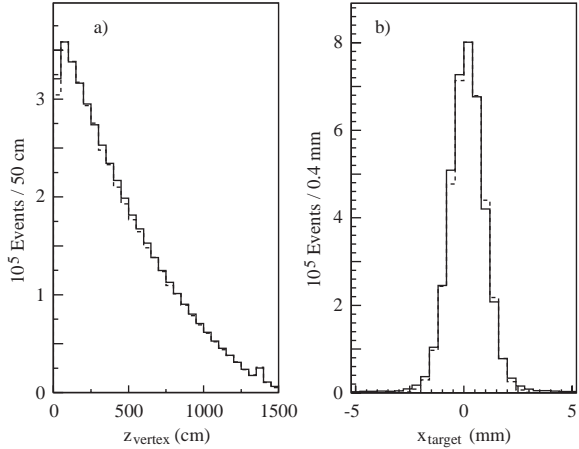}}
\caption{Comparison of normalization-mode Monte Carlo (dashed histogram) and data (solid histogram) distributions in a) $z_{\rm vertex}$ and b) $x_{\rm target}$.}
\label{fig:MCmatch}
\end{figure}

The MC simulation was verified by comparing the simulated \casdec\ events with
data. Distributions of the reconstructed $\Xi^-$ production point, momentum, 
decay vertex, and 
daughters' spatial positions downstream of
the analyzing magnets, for MC events and data, were compared and found to
match well (see Fig.~\ref{fig:MCmatch}). The $\Xi^-$'s 
were generated at the target in
identical fashion for both the signal and normalizing modes. Since a theoretical calculation of the \pmmdec\ decay distribution is not available, it was simulated according to 
three-body phase space. 
The spectrometer
acceptances for the signal and normalizing modes (for $\Xi^-$ hyperons emerging from the downstream  collimator aperture) were estimated to be
9.3\% and 27.4\% respectively, and the respective selection efficiencies (including detection and trackfinding efficiencies)
were 31.7\% and 81.1\%~\cite{decay-model}. 
Table~\ref{tbl-cuteff} shows the effect on MC and data events (in the mass range 1.315--1.330\,GeV/$c^2$) as the selection requirements are imposed.

\begin{table}
\caption{Effect of selection requirements \label{tbl-cuteff}}
\centering
\begin{tabular}{lcccc}
\hline\hline
&  &\multicolumn{3}{c}{\% surviving in}\\
\raisebox{1.5ex}[0pt]{Requirement} & ~ & MC calculation& ~ & data \\
\hline
$\Xi^-$ momentum && 82.6 && 89.9\\
$z_{\rm vertex}$ && 73.6 && 58.7\\
Single-vertex criteria && 59.1 && 8.43\\
$x_{\rm target}, y_{\rm target}$   && 59.1 && 7.29\\ 
$K$ veto && 53.7 && 0.67\\
$\Xi$ and $\Lambda$ vetoes&& 34.9 && 0.15\\
Proton momentum fraction && 31.7 && 0.02\\
\hline\hline
\end{tabular}
\end{table}

No signal was observed, and our result is dominated by statistical uncertainty. 
Nonetheless, we carried out studies of
possible systematic effects. The largest systematic uncertainty, contributing less than $\pm6.4$\%, was due to variations in relative acceptance  between the signal and normalizing
modes due to fluctuations in the position of the beam. The uncertainty due to imperfections in the MC simulation of the signal and normalizing modes was estimated by varying the parameters in the $\Xi^-$ production model; the resulting rms variation  in relative acceptance was found to be less than $\pm 3.1\%$.
Variations in muon-detector efficiencies were studied using data and the systematic
effect on the muon-detection efficiency estimated at
$\pm1.4\%$. 
There is also a $\pm0.78$\% contribution to the normalization uncertainty due to our imperfect knowledge of $\mathcal{B}(\Lambda\to p\pi^-)=(63.9\pm0.5)\%$~\cite{PDG}. (The contribution due to the uncertainty of $\mathcal{B}(\Xi^-\to\Lambda\pi^-)$ is negligible~\cite{PDG}.) The
combined systematic uncertainty in our measurement is thus $\pm7.2$\%. 
To derive the 90\%-C.L. upper limit on the signal branching ratio, we used a Monte Carlo simulation of a large sample of hypothetical experiments that took into account the Poisson fluctuation in the number of signal events observed along with the uncertainty of the background estimate and the uncertainty of the normalizing factor (both treated as Gaussian-distributed)~\cite{Poisson-MC}. The resulting upper limit was $N_{\rm sig}<4.05$ events.

The signal-mode branching fraction is thus
\begin{eqnarray}
\mathcal{B}(\Xi^-&\to& p\mu^-\mu^-) =
\frac{N_{\rm sig}}{N_{\rm norm}}\times \frac{A_{\rm norm}}{A_{\rm sig}}\times \frac{\epsilon_{\rm norm}}{\epsilon_{\rm sig}}\qquad\qquad \nonumber \\
&~&\times \mathcal{B}(\Xi^-\to\Lambda\pi^-)\times\mathcal{B}(\Lambda\to p\pi^-) \\[2mm]
&<&\frac{4.05}{4.92\times10^8}\times \frac{0.274}{0.093}\times \frac{0.811}{0.317} \nonumber 
\times 0.99887 \times 0.639 \nonumber \\[2mm]
&<& 4.0\times10^{-8} {\rm ~at~90\%~confidence}.
\end{eqnarray}
Here, $N$ denotes the number of events observed 
and $A$ and $\epsilon$ are the acceptance and efficiency, with subscripts sig designating the signal mode $\Xi^-\to p\mu^-\mu^-$ and norm the normalizing mode $\Xi^-\to\Lambda\pi^-,\Lambda\to p\pi^-$.

In summary, based on
data from the 1997 run of HyperCP, we see no signal for the lepton-number-violating decay {\pmmdec}. We set an upper limit on the branching ratio
${\mathcal{B}(\Xi^- \rightarrow p \mu^- \mu^-)} < 4.0 \times 10^{-8}$
 at the 90\% confidence level. Our measurement improves upon the
existing limit by four orders of magnitude.

\begin{acknowledgments}
We are indebted to the Fermilab staff for their hard work and dedication and to R. Shrock for valuable discussions. This work was supported by the U.S. Dept.\ of Energy and the National Science Council of Taiwan, R.O.C\@. D.M.K. acknowledges support from the U.K. Particle Physics and Astronomy Research Council and the hospitality of Imperial College London while this paper was in preparation. E.C.D. and K.S.N. were partially supported by the Institute for Nuclear and Particle Physics of the University of Virginia. K.B.L. was partially supported by the Miller Institute for Basic Research in Science. 

\end{acknowledgments}


\begin{thebibliography}{99}

\bibitem{thbackground} L. Wolfenstein, in {\em Proc. XIth Int.\ Conf.\ on Neutrino Physics and Astrophysics}, ed.\ K. Kleinknecht and E. A. Paschos (World Scientific, Singapore, 1984), p.~730; A.~Zee, Phys.\ Lett.\ B \textbf{93}, 389 (1980), A. Zee {\em ibid.}\  {\bf 95}, 461(E) (1980); P. Langacker and D. London, Phys.\ Rev.\ D {\bf 38}, 907 (1988);
E. Witten, Nucl.\ Phys.\ Proc.\ Supp.\ \textbf{91}, 3 (2001); B. Kayser, in {\em Neutrino Mass}, ed.\ G. Altarelli and K. Winter, Springer Tracts in 
Modern Physics Vol.\ 190 (Springer-Verlag, Heidelberg, 2003), p.~1.

\bibitem{neutosc} 
Y. Fukuda {\it et al.}, 
Phys.\ Rev.\ Lett.\ 
{\bf 82}, 2644 (1999); Q. R. Ahmad  {\it et al.},   
{\em ibid.}\ {\bf 89}, 011302 (2002); K. Eguchi {\it et al.}, 
{\em ibid.}\ {\bf 90}, 021802 (2003). 

\bibitem{lnv-models} S.~Weinberg, Phys.\ Rev.\ Lett.\ \textbf{43}, 1566 (1979); R.~Barbieri, J.~Ellis, and
M.~K.~Gaillard, Phys.\ Lett.\ B \textbf{90}, 249 (1980); M.~Gell-Mann, P.~Ramond, and  R.~Slansky, in \textit{Supergravity}, ed.\ P.~Van Nieuwenhuizen and D. Freedman (North Holland, Amsterdam, 1979); T.~Yanagida, Prog.\ Theor.\ Phys.\ {\bf 64}, 1103 (1980); 
J. L. Chkareuli and C. D. Frogatt, Phys.\ Lett.\ B \textbf {484}, 87 (2000).

\bibitem{lnv-maj} J.~Schechter, J.~W.~F.~Valle, Phys.\ Rev.\ D \textbf{25}, 2951 (1982); see also C.~Barbero,
G.~Lopez~Castro,  and A.~Mariano, Phys.\ Lett.\ B {\bf 566}, 98 (2003).

\bibitem{dbdec:heidelmoscow} L. Baudis {\it et al.\ }, Phys.\ Rev.\ Lett.\ \textbf{83}, 41 (1999); see also F.
T. Avignone III, C. E. Aalseth,  and R. L. Brodzinski, {\it ibid.\ }{\bf 85}, 465 (2000).

\bibitem{littenberg2} L.~Littenberg and R.~Shrock, Phys.\ Lett.\ B {\bf 491}, 285 (2000). 

\bibitem{shrock} While the experimental limit presented here is of interest independently of any model, $\mathcal{B}(\Xi^-\to p \mu^-\mu^-)$ has been estimated in the $R$-parity-violating SUSY model of \protect\cite{littenberg2} to be $\sim 
10^{-20}$, although other physics could possibly enhance it  (R. Shrock, private communication).

\bibitem{littenberg} L.~S.~Littenberg and R.~E.~Shrock, Phys.\ Rev.\ D\ \textbf{46}, 892 (1992).

\bibitem{PDG} S. Eidelman {\it et al.}, Phys.\ Lett.\ B {\bf 592}, 1 (2004).

\bibitem{bnl74} N.~Yeh {\textit{et al.}}, Phys.\ Rev.\ D \textbf{10}, 3545 (1974).

\bibitem{hypercp-kmumu} H.~K.~Park \textit{et al.}, Phys.\ Rev.\ Lett.\ \textbf{88}, 111801 (2002); see also R.~A.~Burnstein  \textit{et al.}, hep-ex/0405034 (2004), to appear in Nucl.\ Instrum.\ Meth.

\bibitem{decay-model} The signal-mode acceptance times efficiency ($A_{sig}\epsilon_{sig}
=2.9\%$) is given for the assumed uniform-phase-space decay model, as is standard practice in the absence of a theoretical model; see, e.g., A. Alavi-Harati {\it et al.}, Phys.\ Rev.\ Lett.\ {\bf 93} 021805 (2004) and B. Aubert {\it et al.}, Phys.\ Rev.\ Lett.\ {\bf 92}, 121801 (2004). In five bins of dimuon mass from 0.2 to 0.4\,GeV/$c^2$, the $A\times\epsilon$ values decrease monotonically from 10.3\% at low mass to 1.9\% at high mass.

\bibitem{Poisson-MC} R.~D.~Cousins and
V.~L. Highland, Nucl.\ Instrum.\ Methods A {\bf 320}, 331 (1992); J.~Conrad, O.~Botner, A.~Hallgren, and
C.~Perez~de~los~Heros, Phys.\ Rev.\ D {\bf 67}, 012002 (2003). 

\end{thebibliography}
\end{document}